\documentclass[cits]{PoS}

\title{QCD in terms of gauge-invariant dynamical variables}

\ShortTitle{QCD in terms of gauge-invariant dynamical variables}

\author{\speaker{Hans-Peter Pavel}
\\
        Institut f\"ur Kernphysik, TU Darmstadt, D-64289 Darmstadt, Germany\\ Bogoliubov Laboratory of Theoretical Physics, JINR Dubna, Russia\\
        E-mail: \email{hans-peter.pavel@physik.tu-darmstadt.de}}

\abstract{For a complete description of the physical properties of low-energy QCD,
it might be advantageous to first reformulate QCD in terms of gauge-invariant
dynamical variables, before applying any approximation schemes.
Using a canonical transformation of the dynamical variables, which Abelianises the non-Abelian Gauss-law constraints
to be implemented, such a reformulation can be achieved for QCD.
The exact implementation of the Gauss laws reduces the colored spin-1 gluons and spin-1/2 quarks to unconstrained colorless
spin-0, spin-1, spin-2 and spin-3 glueball fields and colorless Rarita-Schwinger fields respectively.
The obtained physical Hamiltonian can then be rewritten into a
form, which separates the rotational from the scalar degrees of freedom,
and admits a systematic strong-coupling expansion in powers of $\lambda=g^{-2/3}$,
equivalent to an expansion in the number of spatial derivatives.
The leading-order term in this expansion corresponds to non-interacting hybrid-glueballs, whose
low-lying masses can be calculated with high accuracy by solving the Schr\"odinger-equation
of the Dirac-Yang-Mills quantum mechanics of spatially constant physical fields (at the moment only for the 2-color case).
Due to the presence of classical zero-energy valleys of the chromomagnetic potential
for two arbitrarily large classical glueball fields (the unconstrained analogs 
of the well-known constant Abelian fields), practically 
all glueball excitation energy is expected to go into the increase of the strengths of these two fields.
Higher-order terms in $\lambda$  lead to interactions between the hybrid-glueballs and can be taken into account systematically
using perturbation theory in $\lambda$.}

\FullConference{Xth Quark Confinement and the Hadron Spectrum\\
                 8-12 October 2012\\
                 TUM Campus Garching, Munich, Germany}

\begin{document}
%%%%%%%%%%%%%%%%%%%%%%%%%%%
\section{Introduction}

The QCD action
\begin{eqnarray}
{\cal S} [A, \psi,\overline{\psi}] &  = &
\int d^4x\left[ - \frac{1}{4} F^a_{\mu\nu} F^{a\mu \nu}+\overline{\psi}\left(i\gamma^\mu D_\mu-m\right) \psi\right]
\end{eqnarray}
is invariant under the $SU(3)$ gauge transformations
$U[\omega(x)]\equiv \exp(i\omega_a\tau_a/2)$
\begin{eqnarray}
\label{tr}
\psi^{\omega}(x) = U[\omega(x)]\ \psi(x),\quad\quad
A_{a\mu}^{\omega}(x) \tau_a/2   =
U[\omega(x)] \left(A_{a\mu}(x) \tau_a/2 +{i\over g}\partial_\mu \right) U^{-1}[\omega(x)].
\end{eqnarray}
Introducing the chromoelectric
$E_{i}^{a}\equiv F^a_{i 0}$ and chromomagnetic  $B_{i}^{a} \equiv\frac{1}{2}\epsilon_{ijk}F^a_{j k}$  and noting that 
the momenta conjugate to the spatial $A_{ai}$ are $\Pi_{ai}=-E_{ai}$, one obtains the canonical Hamiltonian
\begin{eqnarray}
\!\!\!\!H_C\!\!&=&\!\!\!\!\int\!\!\! d^3x\Bigg[\!{1\over 2}E_{ai}^2+{1\over 2} B_{ai}^2(A) - g A_{ai}\, j_{i a}(\psi)
   + \overline{\psi}\left(\gamma_i\partial_i+m\right) \psi
   - g A_{a0} \left( D_i(A)_{ab} E_{bi} - \rho_a(\psi)\right)\!\!\Bigg]\!,
\end{eqnarray}
with the covariant derivative
$~D_i(A)_{ab}\equiv\delta_{ab}\partial_i-g f_{abc}A_{ci}~$ in the adjoint representation.

Exploiting the {\it time dependence of the gauge transformations} (\ref{tr}) to put (see e.g. \cite{Christ and Lee})
\begin{equation}
A_{a0} = 0~,\quad\quad a=1,..,8 \quad\quad ({\rm Weyl\ gauge}),
\end{equation}
and quantising the dynamical variables $A_{ai}$, $-E_{ai}$, $\psi_{\alpha r}$ and $\psi^*_{\alpha r}$
in the Schr\"odinger functional approach
 by imposing equal-time (anti-) commutation relations (CR) , e.g.  $-E_{ai} = -i\partial/\partial A_{ai}$,
the physical states $\Phi$ have to satisfy both the Schr\"odinger equation and the Gauss laws 
\begin{eqnarray}
\label{constrH}
H\Phi &=& \int d^3x
 \left[{1\over 2} E_{ai}^2+{1\over 2}B_{ai}^2[A]
- A_{ai}\, j_{i a}(\psi)+ \overline{\psi}\left(\gamma_i\partial_i+m\right) \psi\right]\Phi=E\Phi~,
\quad\quad \\
G_a(x)\Phi &=& \left[D_i(A)_{ab} E_{bi}
-\rho_a(\psi)\right]\Phi=0~,
\quad a=1,..,8~.
\label{constrG}
\end{eqnarray}
The Gauss law operators $G_a$ are the generators of the residual {\it time independent gauge transformations} in (\ref{tr}),
satisfying $[G_a(x),H]=0$ and $[G_a(x),G_b(y)]=if_{abc}G_c(x)\delta(x-y)$.

\noindent
Furthermore, $H$ commutes with the angular momentum operators  
\begin{eqnarray}
\label{constrainedJ}
 J_i  = \int d^3x\left[  -\epsilon_{ijk}A_{aj} E_{ak} + \Sigma_i(\psi)+ {\rm orbital \ parts}\right] ~,
\quad i=1,2,3~.
\end{eqnarray}
The matrix element of an operator $O$ is given in the {\it Cartesian} form
\begin{eqnarray}
\langle \Phi'| O|\Phi\rangle\
\propto
\int dA\ d\overline{\psi}\ d\psi\
\Phi'^*(A,\overline{\psi},\psi)\, O\, \Phi(A,\overline{\psi},\psi)~.
\end{eqnarray}

The  spectrum of Equ.(\ref{constrH})-(\ref{constrG}) for the case of Yang-Mills quantum mechanics of spatially constant gluon fields, 
has been found in \cite{Luescher and Muenster} for $SU(2)$ 
and in \cite{Weisz} for $SU(3)$, in the context of a weak coupling expansion in $g^{2/3}$, using the variational approach with gauge-invariant wave-functionals automatically satisfying (\ref{constrG}). 
The corresponding unconstrained approach, 
a description in terms of gauge-invariant dynamical variables via an exact implementation of the Gaws laws, 
has been considered by many authors
(o.a. \cite{Christ and Lee},\cite{Kihlberg and Marnelius}-\cite{pavel2012}, and references therein) 
to obtain a non-perturbative description of QCD at low energy, as an alternative to lattice QCD. 

I shall first discuss in Section 2 the unphysical, but technically much simpler case of 2-colors, and then show in Section 3
how the results can be generalised to $SU(3)$.

%%%%%%%%%%%%%%%%%%%%%%%%%%
\section{Unconstrained Hamiltonian formulation of 2-color QCD}

%%%%%%%%%%%%%%%%%%%%%%%%%%
\subsection{Canonical transformation to adapted coordinates}

Point transformation from the  $A_{ai},\psi_\alpha$ to a new set of adapted coordinates,
the  3 angles $ q_j$ of an orthogonal matrix $O(q)$, 
the  6 elements of a pos. definite symmetric $3\times 3$ matrix $S$, and  new  $\psi^\prime_\beta$
\begin{eqnarray}
\label{polardecomp}
A_{ai} \left(q, S \right) =
O_{ak}\left(q\right) S_{ki}
- {1\over 2g}\epsilon_{abc} \left( O\left(q\right)
\partial_i O^T\left(q\right)\right)_{bc}\,,\quad
\psi_\alpha\left(q, \psi^\prime \right)= U_{\alpha\beta}\left( q \right) \psi^\prime_{\beta}~, 
\end{eqnarray}
where the orthogonal  $ O(q) $  and the unitary   $U\left( q \right)$ are related  via
$O_{ab}(q)={1\over 2}\mbox{Tr}\left(U^{-1}(q)\tau_a U(q)\tau_b\right)$.
Equ. (\ref{polardecomp}) is the generalisation of the (unique) polar decomposition of $A$ and corresponds to 
\begin{eqnarray}
\label{symgaugeSU2}
\quad\quad\chi_i(A)=\epsilon_{ijk}A_{jk}=0\quad(\rm{"symmetric}\ \rm{gauge"}).
\end{eqnarray}
Preserving the CR, we obtain the old canonical momenta in terms of the new variables
\begin{eqnarray}
-E_{ai}(q,S,p,P)=O_{ak}\left(q\right)\left[ P_{ki}+\epsilon_{kil}
{^{\ast}\! D}^{-1}_{ls}(S)\left(\Omega^{-1}_{sj}(q) p_{j}
+\rho_s(\psi^{\prime})
+D_n(S)_{sm} P_{mn}\right)\right].
\end{eqnarray}
In terms of the new canonical variables the Gauss law constraints are Abelianised,
\begin{eqnarray}
\label{unconstrG}
  G_a\Phi
\equiv  O_{ak}(q)\Omega^{-1}_{ki}(q)  p_i\Phi =0\quad
\Leftrightarrow\quad\frac{\delta }{\delta q_i}\Phi=0\quad (\rm{Abelianisation}),
\end{eqnarray}
and the angular momenta become
\begin{eqnarray}
\label{unconstrJ}
 J_i  = \int d^3x\left[  -2\epsilon_{ijk}S_{mj} P_{mk} + 
\Sigma_i(\psi^{\prime})+ \rho_i(\psi^{\prime})+ {\rm orbital \ parts}\right].
\end{eqnarray}
Equ.(\ref{unconstrG}) identifies the $q_i$ with the gauge angles and $S$ and $\psi^{\prime}$ as the physical fields.
Furthermore, from Equ.(\ref{unconstrJ}) follows that the $S$ are {\it colorless} 
spin-0 and spin-2 glueball fields, and the $\psi^{\prime}$ {\it colorless} reduced quark fields of spin-0 and spin-1.
Hence the gauge reduction corresponds to the conversion {\it "color $\rightarrow$ spin"}. The obtained unusual
spin-statistics relation is specific to SU(2).
 
%%%%%%%%%%%%%%%%%%%%%%%%%%
\subsection{Physical quantum Hamiltonian}

According to the general scheme \cite{Christ and Lee}, the {\it correctly ordered physical quantum Hamiltonian} in terms of the
physical variables $S_{ik}({\mathbf{x}})$  and the canonically conjugate 
$P_{ik}({\mathbf{x}})\equiv -i\delta /\delta S_{ik}({\mathbf{x}})$  reads \cite{pavel2010}
\begin{eqnarray}
H(S,P)\!\!\!\!&=&\!\!\!\! {1\over 2}{\cal J}^{-1}\!\!\int d^3{\mathbf{x}}\ P_{ai}\ {\cal J}P_{ai}
     +{1\over 2}\int d^3{\mathbf{x}}\left[B_{ai}^2(S)- S_{ai}\, j_{i a}(\psi^\prime) + \overline{\psi}^\prime\left(\gamma_i\partial_i+m\right) \psi^\prime\right]\nonumber\\
      & & -{\cal J}^{-1}\!\!\int d^3{\mathbf{x}}\int d^3{\mathbf{y}}
      \Big\{\Big(D_i(S)_{ma}P_{im}+\rho_a(\psi^\prime)\Big)({\mathbf{x}}){\cal J}\
        \nonumber\\
      &&\quad\quad\quad\quad\quad\quad\quad\quad\quad\quad
            \langle {\mathbf{x}}\ a|^{\ast}\! D^{-2}(S)|{\mathbf{y}}\ b\rangle\ \
             \Big(D_j(S)_{bn}P_{nj}+\rho_b(\psi^\prime)\Big)({\mathbf{y}})\Big\},
\end{eqnarray}
%with the covariant derivative
%\begin{equation}
%$~D_i(S)_{kl}\equiv\delta_{kl}\partial_i-g\epsilon_{klm}S_{mi}~$,\\
%\end{equation}
with the Faddeev-Popov (FP) operator 
\begin{eqnarray}
^{\ast}\! D_{kl}(S)\equiv \epsilon_{kmi}D_i(S)_{ml}=\epsilon_{kli}\partial_i-g (S_{kl}-\delta_{kl} {\rm tr} S),
\end{eqnarray}
%\gamma_{kl}(S)~,
%   \quad\quad \gamma_{kl}(S)\equiv S_{kl}-\delta_{kl} {\rm tr} S
%\nonumber\end{eqnarray}
and the Jacobian
${\cal J}\equiv\det |^{\ast}\! D| $.
The matrix element of a physical operator O is given by
\begin{eqnarray}
\langle \Psi'| O|\Psi\rangle\
\propto
\int_{\rm S\ pos. def.}\int_{\overline{\psi}^\prime\!,\psi^\prime } 
 \prod_{\mathbf{x}}\Big[dS({\mathbf{x}})d \overline{\psi}^\prime\! ({\mathbf{x}})d \psi^\prime\!( {\mathbf{x}})\Big]
{\cal J}\Psi'^*[S,\overline{\psi}^\prime,\psi^\prime] O\Psi[S,\overline{\psi}^\prime,\psi^\prime].
\end{eqnarray}
 {\it The inverse of the FP operator and hence the physical Hamiltonian
can be expanded in the number of spatial derivatives, equivalent to a strong coupling expansion in $\lambda=g^{-2/3}$}.

%%%%%%%%%%%%%%%%%%%%%%%%%%%%%%%
\subsection{Coarse-graining and strong coupling expansion of the physical Hamiltonian in $\lambda\!=\!g^{-2/3}$}

Introducing an UV cutoff $a$ by considering an infinite spatial lattice of granulas
$G({\mathbf{n}},a)$ at  ${\mathbf{x}}=a {\mathbf{n}}$  $({\mathbf{n}}\in Z^3)$
and averaged variables
\begin{eqnarray}
\label{average}
S({\mathbf{n}}) :=  \frac{1}{a^3}\int_{G({\mathbf{n}},a)} d{\mathbf{x}}\ S({\mathbf{x}})~,
\end{eqnarray}
and discretised spatial derivatives, the expansion of the Hamiltonian in $\lambda=g^{-2/3}$ can be written
\begin{eqnarray}
H =\frac{g^{2/3}}{a}\left[{\cal H}_0+\lambda \sum_\alpha {\cal V}^{(\partial)}_\alpha
                             +\lambda^2 \left(\sum_\beta {\cal V}^{(\Delta)}_\beta
                             +\sum_\gamma{\cal V}^{(\partial\partial\neq\Delta)}_\gamma\right)
                             + {\mathcal{O}}(\lambda^3)\right].
\end{eqnarray}
The "free" Hamiltonian
$ H_0 = (g^{2/3}/a){\cal H}_0+{H}_m
=\sum_{\mathbf{n}} H_0^{QM}({\mathbf{n}}) $
is the sum of the Hamiltonians of Dirac-Yang-Mills quantum mechanics of constant
fields in each box, and the interaction terms 
$ {\cal V}^{(\partial)}, {\cal V}^{(\Delta)},..$ lead to interactions between the granulas.

%%%%%%%%%%%%%%%%%%%%%%%%%%%%%%%%%
\subsection{Zeroth-order: Dirac-Yang-Mills quantum mechanics of spatially constant fields}

\noindent
Transforming to the intrinsic system of the symmetric tensor $S$, with Jacobian $\sin\beta\prod_{i<j}(\phi_i-\phi_j)$,
\begin{eqnarray}
S  =  R^{T}(\alpha,\beta,\gamma)\ \mbox{diag} ( \phi_1 , \phi_2 , \phi_3 )\ R(\alpha,\beta,\gamma),
\quad\quad \psi^{\prime (i)}_{L,R}=R^T_{ij}\widetilde{\psi}^{ (j)}_{L,R},
\quad\quad \psi^{\prime (0)}_{L,R}=\widetilde{\psi}^{ (0)}_{L,R},
\end{eqnarray}
the "free"  Hamiltonian in each box (volume $V$) takes the form \cite{pavel2011}
\begin{eqnarray}
H_0^{QM}&=&{g^{2/3}\over V^{1/3}}\Bigg[{\cal H}^{G}+{\cal H}^{D}+{\cal H}^{C}\Bigg]
+{1\over 2} m\! \left[\!\left(\widetilde{\psi}^{(0)\dag}_L\widetilde{\psi}^{(0)}_R\!
         +\sum_{i=1}^3\widetilde{\psi}^{(i)\dag}_L\widetilde{\psi}^{(i)}_R\right)\! +h.c.\right],
\end{eqnarray}
with the glueball part  ${\cal H}^{G}$, the minimal-coupling ${\cal H}^{D}$, and the Coulomb-potential-type part ${\cal H}^{C}$
\begin{eqnarray}
 {\cal H}^{G}\!\!&=&\!{1\over 2}\!\sum^{\rm
cyclic}_{ijk}\!\! \Bigg(\!\!\! -{\partial^2\over\partial \phi_i^2} -{2\over \phi_i^2-
\phi_j^2}\!\left(\!\phi_i{\partial\over\partial
\phi_i}-\phi_j{\partial\over\partial \phi_j}\right)
 +(\xi_i-\widetilde{J}^Q_i)^2 \!\! {\phi_j^2+\phi_k^2\over (\phi_j^2-\phi_k^2)^2}
+\!  \phi_j^2 \phi_k^2\!\Bigg)\! ,
\\
\label{HphysD}
{\cal H}^{D}\!\!&=&\!\! {1\over 2}(\phi_1+\phi_2+\phi_3)\!
\left(\widetilde{N}^{(0)}_L-\widetilde{N}^{(0)}_R\right)
\!+\!
{1\over 2}\sum^{\rm cyclic}_{ijk}\!\! (\phi_i-(\phi_j+\phi_k))\!
\left(\widetilde{N}^{(i)}_L-\widetilde{N}^{(i)}_R\right),
\\
\label{HphysC}
{\cal H}^{C}\!\!&=&
\!\! \sum^{\rm cyclic}_{ijk}
      {\widetilde{\rho}_i(\xi_i-\widetilde{J}^Q_i+\widetilde{\rho}_i) \over (\phi_j+\phi_k)^2}~,
\end{eqnarray}
\vspace{-0.5cm}
\begin{eqnarray}
{\rm and\ the\ total\ spin }\quad\quad\quad\quad J_i=R_{ij}(\chi)\ \xi_j~,\quad\quad [J_i,H]=0~.
\quad\quad\quad\quad\quad\quad\quad\quad\quad\quad\quad\quad\quad\quad\ \ 
\end{eqnarray}
The matrix elements become
\begin{eqnarray}
\langle\Phi_1|{\cal O} |\Phi_2\rangle \!=\!
\!\!\int\!\!  d\alpha \sin\beta d\beta d\gamma\!\! \int_{0<\phi_1<\phi_2<\phi_3}
\!\!\!\!\!\!\!\!\!\!\!\!\!\!\!\!\!\!\!\!\!\!\!\!\!\!\!\!\!\!\!\!
d\phi_1d\phi_2d\phi_3\ 
(\phi_1^2\!-\!\phi_2^2)(\phi_2^2\!-\!\phi_3^2)(\phi_3^2\!-\!\phi_1^2)\int d\overline{\psi}^\prime d\psi^\prime\Phi_1^* {\cal O}\Phi_2~.\nonumber
\end{eqnarray}
%%%%%%%%%%%%%%%%%%%%%%%%%%%%%%%

The l.h.s. of Fig.1 shows the $0^+$ energy spectrum of the lowest pure-gluon (G) and quark-gluon (QG) cases for one quark-flavor   
 which can be calculated with high accuracy using the variational approach. 
The energies of the quark-gluon ground state and the sigma-antisigma excitation
are lower than that of the lowest pure-gluon state. This is due to a large negative contribution from 
$\langle{\cal H}^{D}\rangle$, in addition to the large positive $\langle{\cal H}^{G}\rangle$,
while  $\langle{\cal H}^{C}\rangle\simeq 0\ $ (see \cite{pavel2011} for details).

Furthermore, as a consequence of the zero-energy valleys $\ "\phi_1\!=\!\phi_2\!=0,\ \ \phi_3\  {\rm arbitrary}"$ of the classical  magnetic  potential $B^2=\phi_2^2\phi_3^2+\phi_3^2\phi_1^2+\phi_1^2\phi_2^2 $, 
practically all glueball excitation-energy results from an increase of expectation value
of the "constant Abelian field" $\phi_3$ as shown for the pure-gluon case on the r.h.s. of Fig.1 (see \cite{pavel2007} for details). 

\begin{figure}
\includegraphics[width=0.55\textwidth]{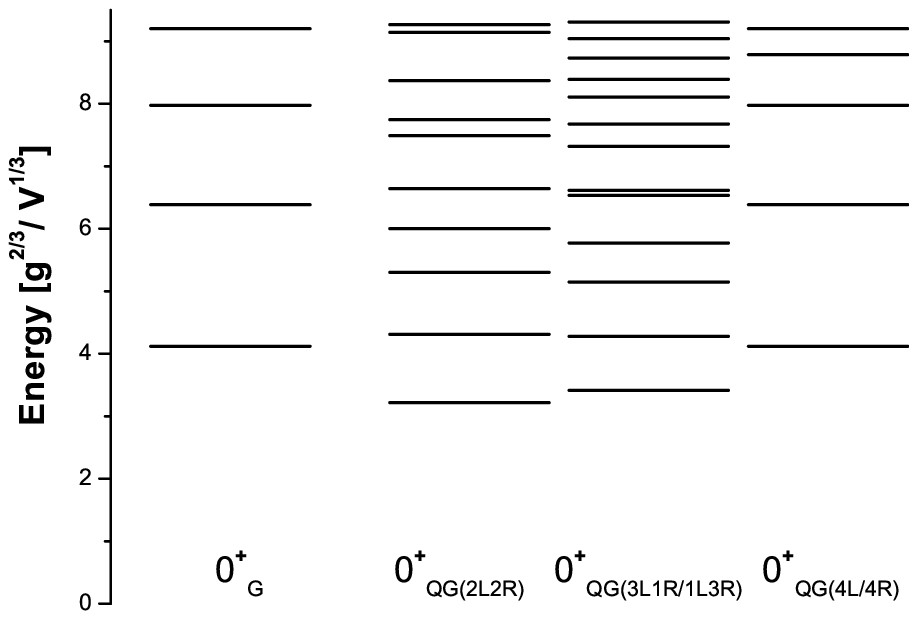}
\includegraphics[width=0.45\textwidth]{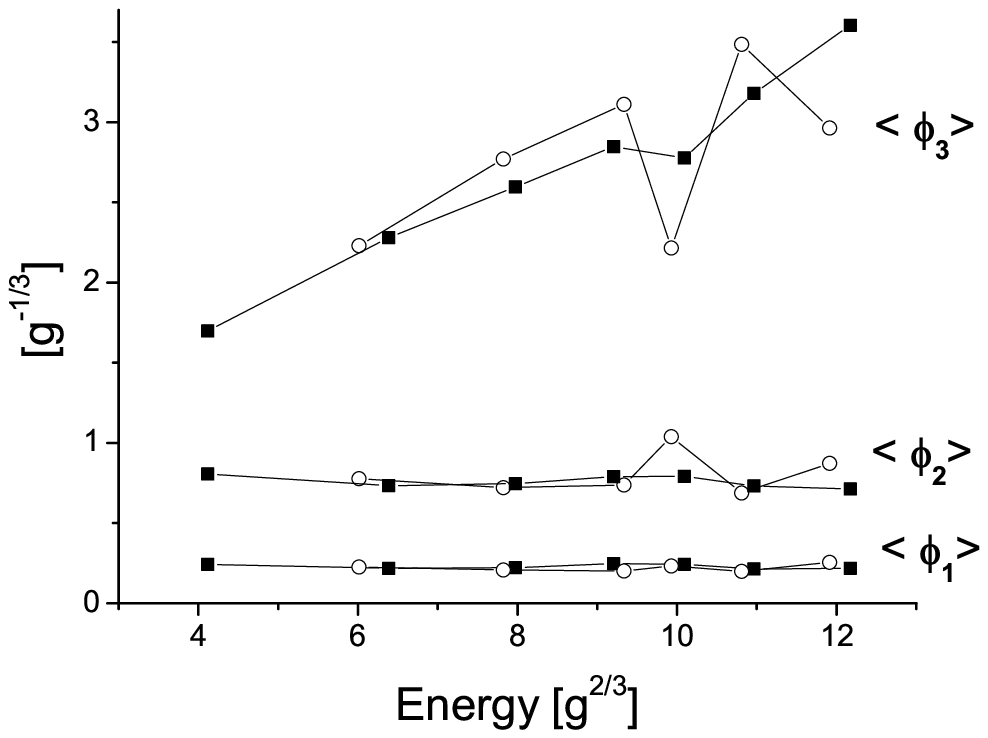}
\caption{L.h.s.: Lowest energy levels for the pure-gluon (G) and the quark-gluon case (QG) for 2-colors and one quark flavor. 
The energies of the quark-gluon ground state and the sigma-antisigma excitation
are lower than that of the lowest pure-gluon state. R.h.s. (for pure-gluon case and setting $V\equiv 1$): 
$\langle \phi_3\rangle$ is raising with increasing excitation,
whereas $\langle \phi_1\rangle$ and $\langle \phi_2\rangle$ are practically constant,
independent of whether spin-0 (dark boxes) or spin-2 states (open circles).}
\label{fig1}
\end{figure}

%%%%%%%%%%%%%%%%%%%%%%%%%%%%%%%%
\subsection{Perturbation theory in $\lambda$ and coupling constant renormalisation in the IR }

Including the interactions ${\cal V}^{(\partial)}, {\cal V}^{(\Delta)}$ using 1st and 2nd order perturbation theory in  $\lambda=g^{-2/3}$ 
give the result \cite{pavel2010} (for pure-gluon case and only including spin-0 fields in a first approximation)
\begin{eqnarray}
E_{\rm vac}^{+}\!&=&\!{\cal N}\frac{g^{2/3}}{a}\Bigg[4.1167+29.894\lambda^2+{\cal O}(\lambda^3)\Bigg],\\
 E_1^{(0)+}(k)-E_{\rm vac}^{+}\! &=&\!
 \left[\ 2.270 + 13.511\lambda^2 + {\cal O}(\lambda^3)\right]\frac{g^{2/3}}{a}
 +0.488 \frac{a}{g^{2/3}} k^2 +{\cal O}((a^2 k^2)^2) ,
\end{eqnarray}
for the energy of the interacting glueball vacuum and the spectrum of the interacting spin-0 glueball.
Lorentz invariance demands $E=\sqrt{M^2+k^2}\simeq M + \frac{1}{2 M}\ k^2 $, which is violated in this 1st approximation
by a factor of $2$. In order to get a Lorentz invariant result,  $J=L+S$ states should be considered including also spin-2 states 
and the general ${\cal V}^{(\partial\partial)}$.

Independence of the physical glueball mass
\begin{equation}
M = \frac{g_0^{2/3}}{a}\left[\mu +c g_0^{-4/3}\right]\nonumber
\end{equation}
of box size $a$, one obtains
\begin{eqnarray}
\gamma(g_0) \equiv a {d\over da} g_0(a) = \frac{3}{2} g_0\frac{\mu+c g_0^{-4/3}} {\mu-c g_0^{-4/3}}
\end{eqnarray}
which  vanishes for 
$g_0 = 0$ (pert.\ fixed\ point) or $g_0^{4/3}=-c/\mu$ (IR\ fixed\ point, if $c<0$). For $c>0$ 
\begin{eqnarray}
{\rm for}\ c>0:\quad\quad g_0^{2/3}(M a) =\frac{M a}{2\mu}
     +\sqrt{\left(\frac{M a}{2\mu}\right)^2
     -\frac{c}{\mu}}~,\quad\quad
     \  a >  a_{c} :=  2\sqrt{c \mu}/M
\end{eqnarray}
My (incomplete) result  $ c_1^{(0)}/\mu_1^{(0)}=5.95 $ suggests, that no IR fixed points exist.
critical coupling $g^2_{0}|_{c}= 14.52$ and $ a_{c} \sim 1.4\ {\rm fm}$ for $ M \sim 1.6\ {\rm GeV}$.

%%%%%%%%%%%%%%%%%%%%%%%%%%%%%%%%
\section{Symmetric gauge for SU(3)}

Using the idea of  {\it minimal embedding} of $su(2)$ in $su(3)$ by Kihlberg and Marnelius \cite{Kihlberg and Marnelius}
\begin{eqnarray}
\label{minimal embedding}
&&\!\!\!\!\!\!\!\!\!\!\!\!
\tau_1 := \lambda_7 ={\small \left(\begin{array}{c c c} 0&0&0\\ 0&0&-i\\ 0&i&0  \end{array}\right)}
\quad
\tau_2 := -\lambda_5 ={\small  \left(\begin{array}{c c c} 0&0&i\\ 0&0&0\\ -i&0&0  \end{array}\right)}
\quad
\tau_3 := \lambda_2 =\!{\small  \left(\!\!\begin{array}{c c c} 0&-i&0\\ i&0&0\\ 0&0&0  \end{array}\!\!\right)}
\nonumber\\
&&\!\!\!\!\!\!\!\!\!\!\!\!
\tau_4 := \lambda_6 = {\small \left(\begin{array}{c c c} 0&0&0\\ 0&0&1\\ 0&1&0  \end{array}\right)}
\quad\quad\!\!
\tau_5 := \lambda_4 = {\small \left(\begin{array}{c c c} 0&0&1\\ 0&0&0\\ 1&0&0  \end{array}\right)}
\quad\quad\
\tau_6 := \lambda_1 =\! {\small \left(\!\!\begin{array}{c c c} 0&1&0\\ 1&0&0\\ 0&0&0  \end{array}\!\!\right)}
\nonumber\\
&&\!\!\!\!\!\!\!\!\!\!\!\!
\tau_7 := \lambda_3 = {\small \left(\begin{array}{c c c} 1&0&0\\ 0&-1&0\\ 0&0&0  \end{array}\right)}
\quad
\tau_8 := \lambda_8 = {1\over \sqrt{3}}{\small \left(\begin{array}{c c c} 1&0&0\\ 0&1&0\\ 0&0&-2  \end{array}\right)}
\end{eqnarray}
such that the corresponding non-trivial non-vanishing structure constants, 
$[{\tau_a\over 2},{\tau_b\over 2}]=  i c_{abc} {\tau_c\over 2}$, have at least one index  $\in \{1,2,3\}$, 
the symmetric gauge, Equ.(\ref{symgaugeSU2}), can be generalised to $SU(3)~$ \cite{Dahmen and Raabe, pavel2012},
\begin{eqnarray}
\chi_a(A)=\sum_{b=1}^8\sum_{i=1}^3 c_{abi} A_{bi}=0~,\quad a=1,...,8\quad\quad ({\rm "symmetric\ gauge"\  for\ SU(3)}).
\end{eqnarray}
%%%%%%%%%%%%%%%%%%%%%%%%%%%
Carrying out the coordinate transformation \cite{pavel2012}
\begin{eqnarray}
A_{ak}\!\left(q_1,..,q_8, \widehat{S} \right)\!
&=&\! O_{a\hat{a}}\left(q \right) \widehat{S} _{\hat{a} k}- {1\over 2g}c_{abc} \left( O\left(q\right)
\partial_k O^T\!\!\left(q\right)\right)_{bc},
\quad
\psi_\alpha\left(q_1,..,q_8, \psi^{RS} \right)= U_{\alpha{\hat{\beta}}}\left( q \right) \psi^{RS}_{\hat{\beta}}\nonumber\\
%\end{equation}
%\begin{eqnarray}
 \widehat{S}_{\hat{a} k }
\equiv {S_{i k }\choose \overline{S}_{A k} }
&=&
\left(\begin{array}{c c c}
                  &  &  \\
 & S_{ik} \ {\rm pos.\ def.}                  & \\
 & & 
\\ \hline
W_0 &
X_3 -W_{3}&
X_2  +W_{2}
\\
X_3  +W_{3}&
W_0 &
X_1  -W_{1}
\\
X_2  -W_{2}&
X_1  +W_{1}&
W_0
\\  
 -{\sqrt{3}\over 2}Y_1  -{1\over 2}W_{1}&
 {\sqrt{3}\over 2}Y_2   -{1\over 2}W_{2}&
  W_{3}
\\
  - {\sqrt{3}\over 2}W_{1} -{1\over 2}Y_1&
  {\sqrt{3}\over 2}W_{2} -{1\over 2}Y_2&
 Y_3
\end{array}\right),\quad  c_{\hat{a}\hat{b} k} \widehat{S}_{\hat{b} k }=0~,
\label{unconstrSU3}
\end{eqnarray}
an unconstrained Hamiltonian formulation of QCD can be obtained.  
The existence and uniqueness of (\ref{unconstrSU3}) can be investigated by
solving the 16 equs.
\begin{eqnarray}
\label{SAequs}
\quad \widehat{S}_{\hat{a} i }\widehat{S}_{\hat{a} j }= A_{ai}A_{aj}\ (6\ {\rm equs.}) 
\quad\wedge\quad 
d_{\hat{a}\hat{b}\hat{c}}\widehat{S}_{\hat{a} i }\widehat{S}_{\hat{b} j }\widehat{S}_{\hat{c} k }= d_{abc}A_{ai}A_{bj}A_{ck}\ (10 \ {\rm equs.}) 
\end{eqnarray}
for the 16 components of $ \widehat{S}$ in terms of 24 given components A.

Analysing the Gauss law operators and the unconstrained angular momentum
operators in terms of the new variables in analogy to the 2-color case, it can be shown that the original constrained 24 colored spin-1 gluon fields $A$
and the 12 colored spin-1/2 quark fields $\psi$ (per flavor)
reduce to {\it 16 physical  colorless spin-0, spin-1, spin-2, and spin-3 glueball fields} (the 16 components of $\widehat{S}$) 
and {\it a colorless spin-3/2 Rarita-Schwinger field} $\psi^{RS}$ (12 components per flavor).
As for the 2-color case, the gauge reduction converts  {\it color $\rightarrow$ spin}, which might have important
consequences for low energy Spin-Physics. In terms of the colorless Rarita-Schwinger fields the  
$\Delta^{++}(3/2)$ could have the spin content $(+3/2,+1/2,-1/2)$ in accordance with the Spin-Statistics-Theorem.

\noindent
Transforming to the intrinsic system of the embedded upper part S of $ \widehat{S}$  (see \cite{pavel2012} for details)
\begin{eqnarray}
S  =  R^{T}(\alpha,\beta,\gamma)\ \mbox{diag} ( \phi_1 , \phi_2 , \phi_3 )\  R(\alpha,\beta,\gamma),
\ \ \wedge\ \ X_i\rightarrow  x_i,\ Y_i\rightarrow y_i ,..,\ \ \wedge\ \ {\psi}^{RS}\rightarrow \widetilde{\psi}^{RS},
\end{eqnarray}
one finds that the magnetic potential $B^2$  has the zero-energy valleys ("constant Abelian fields")
\begin{eqnarray}
B^2=0\ \ :\ \ \phi_3\ {\rm and}\  y_3\  {\rm  arbitrary }\quad \wedge \quad  {\rm all\ others\ zero}
\end{eqnarray}
Hence, practically all glueball excitation-energy should result from an increase of expectation values
of these two "constant Abelian fields", in analogy to $SU(2)$ . 
Furthermore, at the bottom of the valleys the important minimal-coupling-interaction of $\widetilde{\psi}^{RS}$ 
(analogous to (\ref{HphysD})) becomes diagonal
\begin{eqnarray}
 {\cal H}^D_{\rm diag}=
{1\over 2}\widetilde{\psi}_L^{(1,{1\over 2})\dagger}
\left[\left( \phi_3 \lambda_3+y_3 \lambda_8 \right)\otimes \sigma_3 \right]\widetilde{\psi}^{(1,{1\over 2})}_L
 -  {1\over 2}\widetilde{\psi}_R^{({1\over 2},1)\dagger}
\left[ \sigma_3\otimes\left( \phi_3 \lambda_3+y_3 \lambda_8 \right) \right]\widetilde{\psi}^{({1\over 2},1)}_R~.
\end{eqnarray}
Due to the difficulty of the FP-determinant (see \cite{pavel2012}), precise calculations are not possible yet.
%%%%%%%%%%%%%%%%%%%%%%%%%%%%%%%%
Note, however, that in one spatial dimension  the symmetric gauge for SU(3) reduces to 
\begin{eqnarray}
A^{(1d)}={\small
\left(\begin{array}{c c c}
0     & 0 &A_{13}  \\
0 &   0   &A_{23} \\
0 &0 &A_{33} \\ 
0&0&A_{43}\\
0&0&A_{53}\\  
0&0&A_{63}\\
0&0&A_{73}\\
0&0&A_{83} 
\end{array}\right)}
\quad\rightarrow\quad \widehat{S}^{(1d)}= \widehat{S}^{(1d)}_{\rm intrinsic}=
{\small\left(\begin{array}{c c c}
0     & 0 &0  \\
0 &   0   &0 \\
0 &0 & \phi_3\\ 
\hline
0&0&0\\
0&0&0\\  
0&0&0\\
0&0&0\\
0&0& y_3
\end{array}\right)}
\end{eqnarray} which consistently reduces the Equ.(\ref{SAequs}) for given $A_3$ to
\begin{eqnarray}
 \phi_3^2+y_3^2&=&A_{a3}A_{a3}\quad\wedge\quad  \phi_3^2\, y_3-3\, y_3^3= d_{abc}\ A_{a3}A_{b3}A_{c3}
\end{eqnarray}
with 6 solutions  separated by zero-lines of the FP-determinant ("Gribov-horizons"). Exactly one solution exists in the "fundamental domain" 
$\ 0< \phi_3<\infty\ \wedge\  \phi_3/\sqrt{3}<y_3<\infty$, and we can replace
\begin{eqnarray}
\int_{-\infty}^{+\infty} \prod_{a=1}^8 d A_{a3}
&\rightarrow &\int_0^\infty\!\!\!\!\!\! d \phi_3\! \int_{ \phi_3/\sqrt{3}}^\infty d y_3\  \phi_3^2\left( \phi_3^2-3 y_3^2\right)^2
\propto\int_0^\infty\!\!\!\!\!\! r dr\! \int_{\pi/6}^{\pi/2}d \psi \cos^2(3\psi).
\end{eqnarray}
For two spatial dimensions, one can show that (putting in Equ.(\ref{unconstrSU3}) $W_1\equiv X_1,W_2\equiv -X_2$)
\begin{eqnarray}
A^{(2d)}=
{\small\left(\begin{array}{c c c}
A_{11}  &A_{12} & 0 \\
 A_{21}    &A_{22} & 0  \\
A_{31} &A_{32} & 0  \\ 
A_{41}&A_{42} & 0 \\
A_{51}&A_{52} & 0 \\  
A_{61}&A_{62} & 0 \\
A_{71}&A_{72} & 0 \\
A_{81}&A_{82} & 0  
\end{array}\right)}
\rightarrow \widehat{S}^{(2d)}_{\rm intrinsic}=
{\small\left(\begin{array}{c c c}
\phi_1     & 0 &0  \\
0 &   \phi_2    &0 \\
0 &0 & 0\\ 
\hline
0& x_3&0\\
 x_3&0&0\\  
2 x_2&2 x_1&0\\
-{\sqrt{3}\over 2}y_1- {1\over 2}x_1&\ {\sqrt{3}\over 2}y_2+ {1\over 2}x_2 &\ 0\\
-{\sqrt{3}\over 2}y_1+ {1\over 2}x_1&\ -{\sqrt{3}\over 2}y_2+ {1\over 2}x_2&\ 0
\end{array}\right)}
\end{eqnarray}
consistently reduces (\ref{SAequs}) to a system of $7$ equs. for $8$ physical fields (incl. rot.-angle $\gamma$), which, adding as an 8th equ.  
$ (d_{\hat{a}\hat{b}\hat{c}}\widehat{S}_{\hat{b} 1} \widehat{S}_{\hat{c} 2 })^2= (d_{abc}A_{b 1} A_{c 2 })^2$, can be solved numerically 
for randomly generated $A^{(2d)}$, again yielding solutions separated by horizons. Restricting to a fundamental domain
\begin{eqnarray}
\int_{-\infty}^{+\infty} \prod_{a,b=1}^8 d A_{a1} d A_{b2}
\rightarrow\int d\gamma
\int_{0<\phi_1<\phi_2<\infty}\!\!\!\!\!\!\!\!\!\!\!\!\!\!\!\!\!\!\!\!\!\!\!\! d\phi_1 d\phi_2(\phi_1- \phi_2) 
\int_{R_1(\phi_1,\phi_2) }\!\!\!\!\!\!\!\!\!\!\!\!\!\!\!dx_1dx_2 dx_3
\int_{R_2(x_1,x_2,x_3,\phi_1,\phi_2)}\!\!\!\!\!\!\!\!\!\!\!\! dy_1 d y_2\ {\cal J}
\nonumber
\end{eqnarray}
Due to the difficulty of the FP-determinant, I have, however, not yet succeeded in a satisfactory description of the regions $R_1$ and $R_2$.
For the general case of three dimensions, I have found several solutions of the Equ.(\ref{SAequs}) numerically for a randomly generated $A$, 
but to write the corresponding unconstrained integral over a fundamental domain is a difficult, but I think solvable, future task.

%%%%%%%%%%%%%%%%%%%%%%%%%%%%%%%%%%%%%%%%
\section{Conclusions}
Using a canonical transformation of the dynamical variables, which Abelianises the non-Abelian Gauss-law constraints
to be implemented, a reformulation of  QCD in terms of gauge invariant dynamical variables can be achieved.
The exact implementation of the Gauss laws reduces the colored spin-1 gluons and spin-1/2 quarks to unconstrained colorless
spin-0, spin-1, spin-2 and spin-3 glueball fields and colorless Rarita-Schwinger fields respectively.
The obtained physical Hamiltonian admits a systematic strong-coupling expansion in powers of $\lambda=g^{-2/3}$,
equivalent to an expansion in the number of spatial derivatives.
The leading-order term in this expansion corresponds to non-interacting hybrid-glueballs, whose
low-lying masses can be calculated with high accuracy (at the moment only for the unphysical,
but technically much simpler 2-color case) by solving the Schr\"odinger-equation of Dirac-Yang-Mills quantum mechanics of spatially constant fields.
Higher-order terms in $\lambda$  lead to interactions between the hybrid-glueballs and can be taken into account systematically,
using perturbation theory in $\lambda$, allowing for the study of the difficult questions of Lorentz invariance and coupling constant renormalisation in the IR. 
%Furthermore,  the investigation can be extended to flux-tubes (string-tension). 
%The conversion of color to spin in the reduction process might allow for interesting possible insights into low energy Spin-Physics.

%%%%%%%%%%%%%%%%%%%%%%%%%%%%%%%%%%%%%%%%%

\end{document}